\documentclass[apjl]{emulateapj}
\def\gtsima{$\; \buildrel > \over\sim \;$}
\def\gtsim{\lower.5ex\hbox{\gtsima}}

\shorttitle{The Companion of SN Ia}
\shortauthors{Ozaki \& Shigeyama}

\begin{document}

\title{A method to identify the companion stars of type Ia supernovae in young supernova remnants}

\author{Jin Ozaki\altaffilmark{1, }\altaffilmark{2} and Toshikazu Shigeyama\altaffilmark{2}}
\altaffiltext{1}{Department of Astronomy, Graduate School of Science, University of Tokyo, Bunkyo-ku, Tokyo 113-0033, Japan}
\altaffiltext{2}{Research Center for the Early Universe, Graduate School of Science, University of Tokyo, Bunkyo-ku, Tokyo 113-0033, Japan}

\begin{abstract}
We propose a method to identify the companion stars of type Ia supernovae (SNe Ia) in young supernova remnants (SNRs) by recognizing distinct features of absorption lines due to Fe I appearing in the spectrum.  If a sufficient amount of Fe I remains in the ejecta,  Fe I atoms moving toward us absorb photons by transitions from the ground state to imprint broad absorption lines exclusively with the blue-shifted components in the spectrum of the companion star. To investigate the time evolution of column depth of Fe I in the ejecta, we have performed hydrodynamical calculations for SNRs expanding into the uniform ambient media, taking into account collisional ionizations, excitations, and photo-ionizations of heavy elements.  As a result, it is found that the companion star in Tycho's SNR will exhibit observable features in absorption lines due to Fe I at $\lambda\lambda = 371.9935$ nm and 385.9911 nm if a carbon deflagration SN model \citep{w7} is taken. However, these features may disappear by taking another model that emits a few times more intense ionizing photons from the shocked outer layers. To further explore the ionization states in the freely expanding ejecta, we need a reliable model to describe the structure of the outer layers. 
\end{abstract}


\section{Introduction}

Type Ia supernovae (SNe Ia), characterized by no $\rm H_\alpha$ but strong Si lines in the spectra at the maximum brightness, are brighter than most of SNe classified into the other types and exhibit uniform light curves. Thus they are used as a standard candle to measure distances to remote galaxies.

A plausible explosion model for SNe Ia is the accreting white dwarf (WD) model, in which a white dwarf in a binary system accretes material from the companion star, increases its mass, usually up to the Chandrasekhar mass limit (${\rm M_{ch}} \sim 1.4 \, {\rm M}_\odot$), and then explodes \citep[e.g.,][]{Nomoto82}. 
There have been significant progresses in the accreting WD model since \citet{symb} introduced the stellar wind from the WD while it accretes materials from the companion. Their model succeeded in sustaining a stable mass transfer in the progenitor systems of SNe Ia. According to their model, there are two main evolutionary paths leading to SNe Ia, the super soft X-ray source (SSS) channel and the symbiotic channel \citep{symb, sss, rate}. Accordingly their model predicts which companion stars lead to SNe Ia.

The above evolutionary scenario for SNe Ia has not been confirmed by observations, which will require the identification of the companion star that should remain in the vicinity of the explosion site. \citet{tc} argued that their group identified a G2IV star as the companion of Tycho Brahe's supernova remnant (SNR) by measuring the velocities and distances of stars in the vicinity of the center of the SNR. They concluded that this G2IV star was moving much faster than the other neighbor stars and that the distance to this star seemed consistent with the distance to Tycho's SNR. Although \citet{thd} has argued that the observed velocity of Tycho G might correspond to the velocity of stars belonging to the thick disk population, the expected stellar mass in thick disk stars within the cone with 2.87 arcsec radius (which corresponds to the angular distance from the center of Tycho's SNR to the Tycho G star), at 3 kpc from Earth, is only 2 $\times 10^{-3} \, M_\odot$, which makes the thick disk star alternative very unlikely. For this estimate we use the density in the vicinity of Tycho's SNR \citep{sd}.

Although the coincidence of the kinematic characteristics of Tycho G with its being at the position and distance of the SNR appears significant, confirmation by other means would nonetheless be very useful. 

In this paper we propose a direct method to prove that the companion star is located inside the SN ejecta. A hint was dropped by observations for a star called S-M star discovered by \citet{sm} near a type Ia SNR 1006. \cite{uv} proved that the S-M star was not the companion of this supernova by investigating features of Fe II absorption lines in the ultraviolet (UV) spectrum. Very broad wings were observed in both blue and red sides of the absorption lines. The line width was a few thousand km/s much larger than the thermal velocity of stellar atmosphere, which is thought to be $\sim$10 km/s. The broad wings are likely to be formed by Fe II in the ejecta of SNR 1006; photons in the blue wing are absorbed by the matter ejected toward us, and those in the red wing away from us. Thus it was proved that the S-M star is located behind the SNR 1006 ejecta.

When a star is inside the ejecta of SNe Ia, photons emitted from the star are absorbed only by the ejecta moving toward us. Hence the broad wing must be present only in the blue side. The absorption line with only blue wing enables us to identify companion stars of SNe Ia. \citet{uv} used the UV range to observe the S-M star with the Faint Object Spectrograph on the Hubble Space Telescope. However, in addition to difficulties in UV observations from the ground, companion stars on the evolutionary paths suggested by the above mentioned scenario \citep{symb, sss, rate} may not be bright in the UV range. Then we will focus on absorption lines in the visible range.  Furthermore the corresponding transitions need to be from the ground state because most Fe ions  in the ejecta are expected to be in the ground state. Thus only Fe I can produce such absorption lines in SNe Ia ejecta. 

In this paper, we estimate the amount of Fe I in the ejecta of Tycho's SNR by taking account  of collisional processes in non-equilibrium and ionizations by photons emitted from the shocked ejecta, and calculate spectra of a star located at the center of a SNR and discuss whether we can identify the feature of Fe I absorption lines in the spectrum of the companion star.

\begin{table*}[t]
\caption{Fe column densities for SNRs at the age of 430 years.}
\label{tbl:cd}
\begin{center}
\begin{tabular}{c|ccccc}
\hline
\multicolumn{1}{c|}{Density of} & Fe I & Fe II & Fe III & Fe IV & Fe V \\
\multicolumn{1}{c|}{ambient medium} & ($10^{15}\,{\rm cm}^{-2}$) & ($10^{15}\,{\rm cm}^{-2}$) & ($10^{15}\,{\rm cm}^{-2}$) & ($10^{15}\,{\rm cm}^{-2}$) & ($10^{15}\,{\rm cm}^{-2}$) \\
\hline
\multicolumn{1}{c|}{1.0 cm$^{-3}$} & 1.2 & 6.0 & 9.5 & 22 &18 \\
\multicolumn{1}{c|}{0.41 cm$^{-3}$} & 8.9 & 28 & 16 & 8.4 & 1.9 \\
\hline
\end{tabular}
\end{center}
\end{table*}

\section{Formulations}

\subsection{Hydrodynamical Model}
We calculate the hydrodynamical evolution of  a spherically symmetric SN Ia embedded in the ambient medium with a uniform density. The initial structure of freely expanding SN Ia ejecta is taken from W7 in \citet{w7}. The mass of the WD at the time of the SN explosion is 1.37 $\rm M_\odot$. Our model simplifies the original distribution of elements in W7.  That is, the ejecta consist of three layers with different compositions; the innermost layer is consists of pure Fe with the mass of 0.95 $\rm M_\odot$, the middle layer consists of 0.15 $\rm M_\odot$ Si, and the outermost 0.27 $\rm M_\odot$ C/O . The mass fraction of oxygen in this layer is 50 \%. The initial kinetic energy is $1.3\times 10^{51}$ ergs. The density $n_{\rm a}$ of the ambient medium is assumed to be 0.41 $\rm {cm}^{-3}$ \citep{hf} or 1.0 $\rm {cm}^{-3}$ \citep{imn}. These two values refer to two different models of Tycho's SNR.

To obtain the time evolution of physical states in the ejecta, we solve hydrodynamical equations with the Lagrangian coordinate
\begin{eqnarray}
\frac{\partial \tau}{\partial t} - \frac{\partial(r^2u)}{\partial m} & = & 0, \nonumber \\
\frac{\partial u}{\partial t} + r^2 \frac{\partial p}{\partial m} & = & 0, \label{eq:hyd} \\
\frac{\partial E}{\partial t} + \frac{\partial(r^2up)}{\partial m} & = & 0, \nonumber
\end{eqnarray}
where $\tau$ is the specific volume, $u$ the velocity, $E$ the total energy density per unit mass, $m$ the Lagrangian mass coordinate. The internal energy $e$, the mass density $\rho$, and the pressure $p$ are derived from the conserved quantities via
\begin{eqnarray*}
\rho = \frac{1}{\tau}, \; e = E - \frac{u^2}{2}, \; p = (\gamma-1) \rho e,
\end{eqnarray*}
where $\gamma$ is the ratio of specific heats and assumed to be $5/3$ for simplicity. Here we have ignored the radiative cooling, because the associated time-scale is significantly longer than the expansion time-scale of a young SNR like Tycho's SNR \citep[e.g.,][]{sor}. We numerically integrate the above equations (\ref{eq:hyd}) with the PPM \citep{ppm}.

\subsection{Two Temperature Approximation \\ for Shocked Plasma}

In the shocked ejecta, electrons and ions are not immediately relaxed to an equilibrium state because of  their different masses and low densities. Two temperature approximation is used to describe this relaxation process. The time scale for the electron temperature $T_{\rm e}$ to catch up with the ion temperature $T_{\rm i}$ is estimated by \citet{sp}. The time scale $t_{\rm eq}$, which is of the order of $10^4$ years, is longer than the age of Tycho's SNR. Then we must calculate the electron temperature by solving the equation
\begin{equation}
\frac{\partial T_e} {\partial t} = \frac{T_{\rm i} - T_{\rm e}} {t_{\rm eq} ({\rm i},{\rm e})}.
\end{equation}

\subsection{Collisional Processes in the Shocked Region}
In the shocked region, the electron temperature becomes higher than $10^7$ K. Thus collisional excitations and ionizations are activated and dominated. To obtain the population $f(X^j)$ of an element $X$ in the ionization stage $j$, we solve rate equations written as
\begin{equation}
\frac{\partial f(X^j)}{\partial t} = I^{j-1} f(X^{j-1}) - I^j f(X^j),
\end{equation}
where $I^j$ denotes the rate of ionizations out of  the $j$th ion stage. The collisional ionization cross section is calculated using the formula introduced in \citet{lotz}. The excitation-autoionization (E-A) is taken into account for  ions such as Fe$^{3+}$, Fe$^{4+}$, Fe$^{10+}$, Fe$^{11+}$, Fe$^{12+}$, Fe$^{13+}$, Fe$^{14+}$, Fe$^{15+}$, Fe$^{23+}$, Si$^{3+}$, Si$^{11+}$, O$^{5+}$, and C$^{3+}$ \citep{ea}. On the other hand, we can ignore recombination rates in the above equations, because the recombination time scale becomes of the order of $10^5$ years much longer than the age of Tycho's SNR.

Instead of taking into account photo-ionizations in front of the shock, atoms immediately after the shock front in the ejecta are assumed to be automatically ionized to Fe$^{10+}$, Si$^{7+}$, O$^{5+}$, C$^{4+}$ to account for the detection of Si XIII and Fe XVII lines by the XMM-Newton \citep{xro}. This assumption has been made by \citet{imn}. The photo-ionization process is taken into account as a post-process to reduce the computational time.

\subsection{Photon Emission}
A lot of high energy photons are emitted from ions of Si, Fe, O, and C in the shocked ejecta \citep{hf,vink}. These photons ionize the matter in the pre-shocked ejecta.  The number of ionizing photons emitted from ions $i$ per unit volume per unit solid angle per unit time is
\begin{equation}
\epsilon_{\rm {ph}} = \frac{1}{4 \pi} n_{i} n_{\rm e} C_{gj} \; [\rm cm^{-3}\,s^{-1}\,str^{-1}],
\end{equation}
where $n_i$ is the number density of ions $i$, and $C_{gj}$ is the collisional excitation coefficient for the corresponding transition \citep{kato}.

We include ionizing photons emitted from transitions of ions listed in \citet{hf} and \citet{hs}. Strong emission lines from carbon ions ($gf > 10^{-2}$) are taken from \citet{ll1} and \citet{ll2}.

\subsection{Radiative Transfer}
With the source term for ionizing photons discussed in the preceding subsection, we solve the radiative transfer equation for ionizing photons to estimate the ionization states of Fe, Si, O, and C in the pre-shocked ejecta. The intensity $I$ of ionizing photons is obtained by solving the transfer equation,
\begin{equation}\label{rt}
\mu \frac{\partial I}{\partial r} + \frac{1-\mu^2}{r} \frac{\partial I}{\partial \mu} = \epsilon_{\rm {ph}} - \sum_{{\rm el},i} \sigma_{\rm {pi, \, el}} n_{{\rm el} \, i} I,
\end{equation}
where $r$ denotes the distance from the center, $\theta$ the angle between the direction of incident photons and the radial direction, $\mu = \cos\theta$, $\sigma_{\rm {pi,\,el}}$ is the photo-ionization cross section of Fe, Si, O, or C \citep{ver}, and 'el, $i$' represents $(i-1)$ times ionized element el. This equation is integrated along the characteristics to obtain $I(r,\,\theta)$ in the freely expanding ejecta as described in textbooks \citep[e.g.,][]{mih}.
This equation is solved after the optical depth becomes smaller than $\sim 10^2$.

\subsection{Ion Fraction}
The number $f_{\rm ph}$ of photons incident from all directions per unit time is obtained by integrating the intensity $I$ with respect to angle. If the corresponding value for photons that can ionize ions $X^{j}$ is denoted by $f_{{\rm ph},\, X^{j}}$, then the ion fraction $f(X^{j})$ of element $X$ in the $j$th ionization stage evolves with time according to the following equations,
\begin{eqnarray}
\frac{\partial f(X^j)}{\partial t} & = & f(X^{j-1}) (\sigma_{{\rm pi},\,X^{j-1}} f_{{\rm ph},\,X^{j-1}} + n_{\rm e} \langle v \sigma_{\rm col} \rangle^{j-1} ) \nonumber \\
 & & - f(X^j) (\sigma_{{\rm pi},\,X^j} f_{{\rm ph},\,X^j} + n_{\rm e} \langle v \sigma_{\rm col} \rangle^j ) 
\end{eqnarray}
where $\sigma_{{\rm pi},\,X^j}$ is the cross section for photo-ionizing $j$th ions of an element $X$ and $n_{\rm e}\langle v \sigma_{\rm col} \rangle^j$ is the collisional ionization rate.

\section{Results}

\begin{figure}
\begin{center}
\plotone{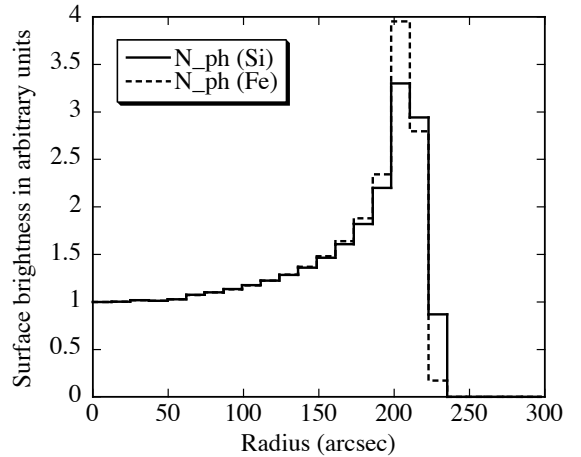}
\caption{Radial profile of X-ray surface brightnesses of Si emission (1670-2000 eV; solid line) and Fe XVII emission line ($\lambda = 15 \, {\rm \AA}$; dashed line) from a model with $n_{\rm a}=1.0$ cm$^{-3}$ at a distance of 2.43 kpc from the earth.}
\label{fig:xrs}
\end{center}
\end{figure}

\subsection{X-ray Surface Brightness}
Comparing the calculated X-ray surface brightness profile with that observed for Tycho's SNR  \citep[e.g.,][]{xro}, one can constrain the density in the ambient medium. Figure \ref{fig:xrs} shows the calculated X-ray surface brightness of Si XIII triplet blend  (in the range of  $1670- 2000$ eV) and Fe XVII line (at $\lambda = 15 \, {\rm \AA}$) emitted from a SNR model with $n_{\rm a}=1.0$ cm$^{-3}$ located at a distance of 2.43 kpc from the earth. The plotted profile has been smoothed by a Gaussian filter of radius 5$^{\prime\prime}$ and binned to a 12.4$^{\prime\prime}$ interval to mimic the observation. The observed radius of the peak is reproduced quite well by this model, while the contrast between the peak brightness and the central brightness is slightly more pronounced in the model. This might be due to a deviation from spherical symmetry in Tycho's SNR, which will lower the peak brightness when plotted assuming spherical symmetry.

\subsection{Column densities of Fe ions}
The column densities of Fe ions thus obtained for SNRs at the age of Tycho's SNR are listed in Table \ref{tbl:cd}.

Comparing with those of \citet{hf}, our results  show that ions tend to populate in lower ionization stages in all calculations. There seem to be two differences in these two calculations. The main difference is in the treatment of carbon ions as a source of ionizing photons. \citet{hf} replaces carbon with oxygen  in calculating the number of ionizing photons, while we do calculate the number of ionizing photons from carbon ions. The other is a difference in hydrodynamical models. We use the W7 model, but they use the CDTG7 model constructed by \citet{Woosley86}. Because the CDTG7 model has a less centrally condensed density profile, photo-ionizing emission from the outer layers are more intense than in the W7 model. Both of these two differences enhance photo-ionizations in \citet{hf} as compared with our calculation. 

\section{Spectra of the Companions}

\subsection{Radiative Transfer}
Some photons emitted in the visible band from the companion star are absorbed by Fe I in the expanding ejecta.  Suppose the companion star emits photons with the frequency $\nu_{\rm s}$ toward an observer. These photons are absorbed by Fe I moving with the velocity $v$, when
\begin{equation}
\nu_{\rm s} = \nu_0\left(1+\frac{v}{c}\right),
\end{equation}
is satisfied as long as $v$ is much smaller than the speed of light $c$. Here $\nu_0$ denotes the frequency of Fe I absorption line in the co-moving frame. Thus the absorbed photons must have a frequency higher than $\nu_0$.

 The optical depth $\tau(\nu_{\rm s})$ of the pre-shocked ejecta for photons with the frequency $\nu_{\rm s}$ is given by  \begin{equation}
\tau(\nu_{\rm s})=\int_{0}^{R}\sigma_{\rm abs} n_{\rm {Fe\,I}}dr.
\end{equation}
Here the radius of the inner shock front is denoted by $R$. This integration can be transformed into the integration with respect to frequency  as
\begin{equation}
\int_{0}^{R}\sigma_{\rm abs} n_{\rm {Fe\,I}}dr = \int^{\nu_{\rm s}(1-R/ct)}_{\nu_{\rm s}}\sigma_{\rm abs} n_{\rm {Fe\,I}}\left(\frac{dr}{dv}\right)\left(\frac{dv}{d\nu}\right)d\nu, \nonumber
\end{equation}
where $\nu$ denotes the photon frequency in the co-moving frame, i.e., $\nu = \nu_{\rm s}(1-v/c)$, $\sigma_{\rm abs}$ the absorption cross section, and $n_{\rm {Fe\,I}}$ the number density of Fe I.

 Since the velocity of the ejecta  (about several thousand km/s) is much larger than the velocity of the thermal motion of particles in the pre-shocked ejecta (less than a ten km/s), applying the Sobolev approximation to the integration above yields,
\begin{equation}
\tau(\nu_{\rm s})=\left\{
\begin{array}{ll}
  \frac{\pi e^2}{m_{\rm e}c}gfn_{\rm Fe\,I} \left(\frac{dr}{dv}\right)\left|\frac{dv}{d\nu}\right|& \mbox{if} \; {\nu_0} < \nu_{\rm s} < {\nu_0(1+R/ct)}, \\
0 & \; \mbox{otherwise}.\end{array}
\right. 
\end{equation}
Here the density $n_{\rm FeI}$ and the derivative $dv/d\nu$ are evaluated at the point where $\nu_0 = \nu_{\rm s}(1-r/ct)$ is satisfied.

We investigate the transfer of photons in the U band absorbed by four transitions of Fe I listed in Table \ref{tbl:ew}.

Though we have assumed that the distribution of Fe is a function of the distance from the center, Fe may be distributed in blobs. We will briefly discuss how these blobs affect the absorption lines discussed in this section. Suppose Fe is distributed in $N$ identical blobs with spherical shapes. These blobs fill a fraction $f$ of the entire volume of the ejecta. Then the column depth of Fe is reduced when 
\begin{equation}
\frac{3}{4}f^{2/3}N^{1/3}<1,
\end{equation}
is satisfied. Otherwise the column depth will not be changed but the shape of the absorption lines will be changed depending on where blobs on the line of sight are located in the ejecta. A 3-dimensional modeling of SNe Ia \citep{Reinecke02} demonstrated that a larger number ($N\gtsim 10$) of bubbles is preferred to accommodate the observed amount of $^{56}$Ni (Fe) and explosion energy. Thus a large $N$ expected from this 3-dimensional modeling is unlikely to reduce the column depth.

Clumpiness in the ejecta was suggested by recent observations for SN 2004dt \citep{Wang04}. Spectropolarimetry of this supernova more than a week before its optical maximum reveals that the Si layer is substantially deformed from spherical symmetry while the same observation shows little polarization signature in O lines. Since Si is not a major source of ionizing photons, the influence to the Fe I lines is insignificant. From the theoretical point of view, however, the clumpy distribution of Fe I can not be ignored \citep[e.g.][]{Reinecke02}. Then we will calculate the spectra from a clumpy distribution of Fe I in the next section though the distribution is rather artificial. To mimic clumpy structures especially in the inner region of the ejecta expected from results of 3-D calculations, the density of the innermost ejecta with velocity $<$2,000 km/s is reduced by a factor of two and the reduced mass is added to the adjacent outer zones with the velocity range of 2,000-3,000 km/s by increasing the densities thereof by a constant factor. Here we assume that the ion fractions are not affected by the clumpiness.

\begin{figure}
\begin{center}
\epsscale{1.0}
\plotone{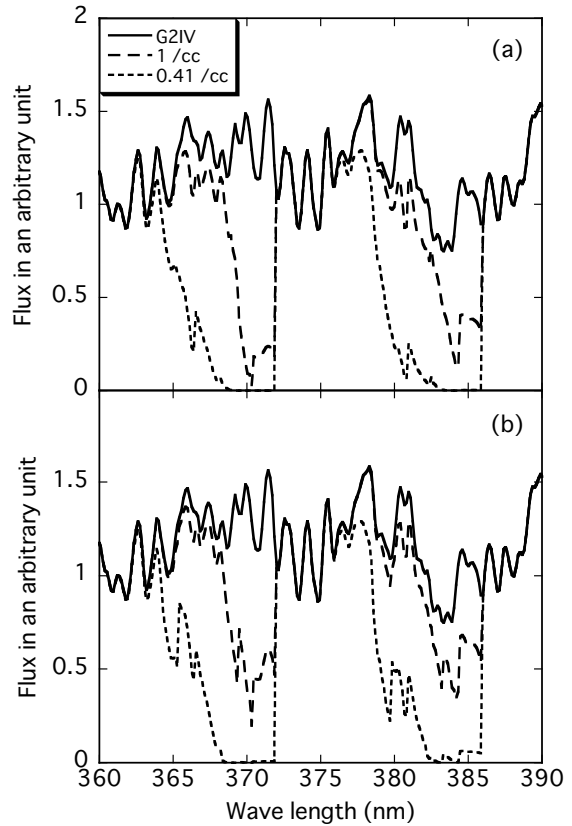}
\caption{(a) Spectra of a G2IV star located at the center of a SNR at the age of Tycho's SNR.  The long dashed line shows the spectrum for $n_{\rm a} = 1.0$ cm$^{-3}$ and the short dashed line for $n_{\rm a} = 0.41$ cm$^{-3}$. The solid line exhibits the original spectrum of the star. (b) the same as (a) but for a star embedded in  clumpy ejecta described in section 4.1.}
\label{fig:g2}
\end{center}
\end{figure}

\subsection{Spectra}
The spectra of a G2IV star located at the center of SNRs embedded in the ambient media with two different densities are  calculated and shown in Figure \ref{fig:g2} (a). This star has the same spectral type as the candidate star \cite{tc} found.

Figure \ref{fig:g2}(b) shows the spectra of the same star in the ejecta in which the distribution of Fe I is affected by the clumping discussed in the previous subsection. It is clear that the clumpy distribution of Fe I changes the shapes of absorption lines: the lower velocity components of the lines are weaken because of the reduced density in the corresponding layers while the addition of mass to the outer layer hardly affects the line shape because a substantial amount of Fe is already ionized there. The equivalent width of each line exhibited in Figure \ref{fig:g2} (a) is listed in Table \ref{tbl:ew}. This particular clumpiness in Fe I distribution  reduces the equivalent width by about 30 \%. The spectra exhibit strong blue wings both for tenuous and dense ambient media. Results of our calculations indicate that Tycho's SNR has a sufficient amount of Fe I to identify the companion, even though a companion star as late as G2 has a very complicated spectrum with many absorption features.

\begin{table}
\caption{Equivalent widths of Fe I lines}
\begin{center}
\begin{tabular}{c|c|c|c}
\hline
\multicolumn{1}{c|}{Fe I Line} &  & \multicolumn{2}{c}{Equivalent widths} \\
\multicolumn{1}{c|}{Wave Length} & $\log(gf)$ & \multicolumn{2}{c}{(nm)}   \\ \cline{3-4}
\multicolumn{1}{c|}{(nm)} &  & $n_{\rm a}=$1.0 cm$^{-3}$ & $n_{\rm a}=$ 0.41 cm$^{-3}$  \\
\hline
\multicolumn{1}{c|}{367.9913} & $-1.599$ & 0.363 & 1.95  \\
\multicolumn{1}{c|}{\textbf{371.9935}} & ${\bf -0.431}$ & \textbf{2.51} & \textbf{4.88}  \\
\multicolumn{1}{c|}{382.4444} & $-1.362$ & 0.633 & 2.86  \\
\multicolumn{1}{c|}{385.9911} & $-0.710$ & 1.94 & 4.59 \\
\hline
\end{tabular}
\end{center}
\label{tbl:ew}
\end{table}

Even when we find a star that exhibits the absorption feature discussed in this paper, there is a chance that a star other than the companion star happens to be inside the ejecta. A star existing in the vicinity of a SN Ia gets a fraction of the explosion energy and is accelerated. If the companion star similar to Tycho G star with the mass of $M_{\rm c} =$ 1.2 $M_\odot$ and the radius of $R_{\rm c} =$ 2 $R_\odot$ is located at the distance of $R_{\rm o} =$ 5.5 $R_\odot$ from the progenitor, the size of the Roche lobe is comparable to the size of the companion. Thus the star in this situation will get the maximum velocity after the explosion. Then the orbital velocity before the explosion is $\sim 160$ km/s. In a 3-D hydrodynamical simulation of SNe Ia by \citet{mari}, they obtain a plausible kick velocity $v\sim 50$ km/s. Therefore, including the orbital velocity, the velocity of the companion star becomes $v_{\rm c}\sim 170$ km/s. If the companion star has been moving away from the explosion site at that speed, the companion star is now at the distance of $<$0.08 pc from the center. The stellar mass density in the neighborhood of Tycho's SNR estimated from a model of the Galaxy  \citep{sd} is less than 0.03 $M_\odot /{\rm pc}^{-3}$. Thus the expected stellar mass inside this volume is only $\sim 10^{-4}\,M_\odot$. Therefore if we find a star showing absorption lines with broad blue wings in the spectrum, it is likely that the star is the companion star.

\section{Conclusions and Discussions}

As a consequence, we have demonstrated that there exhibit very deep absorption lines with unique shapes  in the spectrum of the companion star located at the center of a young SNR such as Tycho's SNR. There are, however, a few factors that might reduce or even erase these distinct absorption features.

First, the number of Fe I in the freely expanding ejecta is very sensitive to the number of ionizing photons emitted from the shocked ejecta. An increase in the number of ionizing photons by a few factors might decrease the number of Fe I by a few orders of magnitude or more. Since the main source of ionizing photons is O in the outer ejecta, the distributions of O and density in the outer ejecta need to be known precisely. Unfortunately, these regions in W7 have some problems to reproduce the observed optical spectra \citep[][and references therein]{Branch98}. Due to this uncertainty in the explosion model, it is not conclusive if the companion star of Tycho's SN will exhibit unique Fe I absorption features discussed in this paper. Nevertheless, it is true that every SN Ia has a period during which the companion star has the distinct absorption features discussed above because the ejecta become cool enough to have plenty of Fe I for a time after the optical brightening.

Second, it is assumed in our calculations that ions in the shocked region are ionized to Fe$^{10+}$, Si$^{7+}$, O$^{5+}$, C$^{4+}$ immediately after the shock passes following the procedure taken by \citet{imn}. Since ions in lower ionization stages are a strong source of ionizing photons, ionization may be more advanced in real young SNRs.

\acknowledgments
This work has been partially supported by Grant-in-Aid for Scientific Research (16540213), from the Ministry of Education, Culture, Sports, Science, and Technology of Japan.

%\clearpage

%\clearpage

%\clearpage

%\clearpage

\end{document}